\documentclass[conference]{IEEEtran}
\IEEEoverridecommandlockouts
\usepackage{cite}
\usepackage{amsmath,amssymb,amsfonts}
\usepackage{algorithm}
\usepackage{graphicx}
\usepackage{textcomp}
\usepackage{xcolor}
\usepackage{amsmath,amssymb,amsfonts}

\usepackage{mathtools}
\usepackage{algorithm}
\usepackage{hyperref}
\hypersetup{
    colorlinks=true,
    linkcolor=red,
    filecolor=magenta,      
    urlcolor=blue,
    citecolor=violet
}%https://www.overleaf.com/project/62a378a8695ed49af7243eb6

\usepackage{xcolor, soul}

\usepackage[noend]{algpseudocode} %this loads the algorithmicx package
\usepackage{graphicx}
\usepackage{tabularx}
\usepackage{multirow}

\usepackage{amsthm}
\usepackage{url}
\usepackage{bm}
\usepackage{tikz} 
\usepackage{wrapfig}
\usepackage{textcomp}
\usepackage{xcolor}
\usepackage{caption}
\usepackage{subcaption}
\usepackage{comment}
\usepackage{mathtools}
\usepackage{xparse}
\usepackage{arydshln}
\usepackage{amssymb}% http://ctan.org/pkg/amssymb
\usepackage{pifont}% http://ctan.org/pkg/pifont

\usepackage{tablefootnote}

\usepackage{color,soul}
\usepackage{lipsum}

\usepackage{todonotes}
\usepackage{footnote}
\usepackage{hyperref}
\usepackage{xcolor}
\usepackage[mathscr]{eucal}
\usepackage{bm}
\usepackage{algpseudocode} 
\usepackage{algorithm}
\usepackage{xifthen}
\usepackage{comment}
\usepackage[export]{adjustbox}
\usepackage{amsmath,amssymb,amsfonts}

\NewDocumentCommand{\vect}{ O{} O{} m }{\mathbf{#3}\ifthenelse{\isempty{#1}}{}{^{(#1)}}\ifthenelse{\isempty{#2}}{}{_{#2}}}

\NewDocumentCommand{\mat}{ O{} O{} m }{\mathbf{#3}\ifthenelse{\isempty{#1}}{}{^{(#1)}}\ifthenelse{\isempty{#2}}{}{_{#2}}}

\NewDocumentCommand{\ten}{ O{} O{} m }{\pmb{\mathscr{#3}}\ifthenelse{\isempty{#1}}{}{^{(#1)}}\ifthenelse{\isempty{#2}}{}{_{#2}}}

\begin{document}

%%
%% The "title" command has an optional parameter,
%% allowing the author to define a "short title" to be used in page headers.
\title{SeNMFk-SPLIT: Large Corpora Topic Modeling by Semantic Non-negative Matrix Factorization  with Automatic Model Selection}

\author{\IEEEauthorblockN{Maksim E. Eren}
\IEEEauthorblockA{\textit{Theoretical Division}\\
\textit{LANL}\\
Los Alamos, U.S\\
maksim@lanl.gov}
\\
\IEEEauthorblockN{Kim \O. Rasmussen}
\IEEEauthorblockA{\textit{Theoretical Division}\\
\textit{LANL}\\
Los Alamos, U.S\\
kor@lanl.gov}
\and
\IEEEauthorblockN{Nick Solovyev}
\IEEEauthorblockA{\textit{Theoretical Division}\\
\textit{LANL}\\
Los Alamos, U.S \\
nks@lanl.gov}
\\
\IEEEauthorblockN{Charles Nicholas}
\IEEEauthorblockA{\textit{CSEE, }\\
\textit{University of Maryland}\\
Baltimore County\\
%Los Alamos, U.S\\
nicholas@umbc.edu}
\and
\IEEEauthorblockN{Manish Bhattarai}
\IEEEauthorblockA{\textit{Theoretical Division}\\
\textit{LANL}\\
Los Alamos, U.S\\
ceodspspectrum@lanl.gov}
\\
\IEEEauthorblockN{Boian S. Alexandrov}
\IEEEauthorblockA{\textit{Theoretical Division}\\
\textit{LANL}\\
Los Alamos, U.S\\
boian@lanl.gov}

}

%%
%% The "author" command and its associated commands are used to define
%% the authors and their affiliations.
%% Of note is the shared affiliation of the first two authors, and the
%% "authornote" and "authornotemark" commands
%% used to denote shared contribution to the research.

%%
%% By default, the full list of authors will be used in the page
%% headers. Often, this list is too long, and will overlap
%% other information printed in the page headers. This command allows
%% the author to define a more concise list
%% of authors' names for this purpose.
%\renewcommand{\shortauthors}{Eren et al.}
\maketitle
%%
%% The abstract is a short summary of the work to be presented in the
%% article.
\begin{abstract}

As the amount of text data continues to grow, topic modeling is serving an important role in understanding the content hidden by the overwhelming quantity of documents. One popular topic modeling approach is non-negative matrix factorization (NMF), an unsupervised machine learning (ML) method. Recently, Semantic NMF with automatic model selection (SeNMFk) has been proposed as a modification to NMF. In addition to heuristically estimating the number of topics, SeNMFk also incorporates the semantic structure of the text. This is performed by jointly factorizing the term frequency-inverse document frequency (TF-IDF) matrix with the co-occurrence/word-context matrix, the values of which represent the number of times two words co-occur in a predetermined window of the text. In this paper, we introduce a novel distributed method, SeNMFk-SPLIT, for semantic topic extraction suitable for large corpora. Contrary to SeNMFk, our method enables the joint factorization of large documents by decomposing the word-context and term-document matrices separately. We demonstrate the capability of SeNMFk-SPLIT by applying it to the entire artificial intelligence (AI) and ML scientific literature uploaded on arXiv.

\end{abstract}

%%
%% The code below is generated by the tool at http://dl.acm.org/ccs.cfm.
%% Please copy and paste the code instead of the example below.
%%

%%
%% Keywords. The author(s) should pick words that accurately describe
%% the work being presented. Separate the keywords with commas.
%%
%% This command processes the author and affiliation and title
%% information and builds the first part of the formatted document.
%\maketitle

\section{Introduction}
\label{sec:introduction}
According to a recent report, 2.9 million research articles are published annually and that number is growing at the rate of 5\% per year \cite{white2021publications}. This substantial expansion of text data, including scientific literature, requires automated document engineering techniques that can handle large-scale data to produce actionable results. One important method for document organization is topic modeling, which maps a collection of documents into themes that summarize the hidden (or latent) features of those documents. By grouping text samples into topics via an unsupervised method, large datasets become easier to understand and process.

In our work, we base our methodology on non-negative matrix factorization (NMF), a popular unsupervised machine learning (ML) method for topic modeling. NMF performs a low-rank approximations of a given term frequency-inverse document frequency (TF-IDF) matrix $\mat{X} \in \rm I\!R_{+}^{m \times n}$ (where $m$ is the number of tokens in the vocabulary and $n$ is the number of documents) by a product of two non-negative factor matrices $\mat{W} \in \rm I\!R_{+}^{m \times k}$ and $\mat{H} \in \rm I\!R_{+}^{k \times n}$, such that $\mat{X}_{ij} \approx \sum^{k}_{s} \mat{W}_{is} \mat{H}_{sj}$, and the low-rank $k\ll n,m$. $\mat{X}$ can be factorized using \textit{multiplicative update}, a minimization algorithm with a non-negativity constraint, that minimizes the non-convex objective $||\mat{X}-\mat{W}\mat{H}||^{2}_{F}$, where $||...||_F$ denotes he Frobenius norm \cite{lee1999learning}. Here the columns of $\mat{W}$ represent the topics and the rows of $\mat{H}$ are the coordinates of the documents in the latent topic space. While NMF and its variants have been successfully applied to various corpora \cite{stanev2021topic, du2017dc, wang2021deep}, classical NMF relies on manual selection of the number of latent topics $k$, which is crucial for identifying meaningful topics. Too small value of $k$ can result on poor topic separation (\textit{under-fitting}), and too large a value of $k$ will result in noisy topics (\textit{over-fitting}). In addition, traditional NMF does not incorporate semantics of the text, which is essential for extracting coherent topics \cite{ailem2017non, salah2018word, shi2018short}.

Semantic NMF with automatic model determination (SeNMFk) is a topic modeling approach that incorporates semantics and estimates the number of latent features $k$ to extract coherent and meaningful topics \cite{9521777}. SeNMFk incorporates semantic information by joint factorization  of the TF-IDF and co-occurrence/word-context matrices. In this paper, we introduce a new topic modeling method, named SeNMFk-SPLIT, that is designed for large-scale data. SeNMFk-SPLIT uses SPLIT where the TF-IDF and word-context matrices are factorized separately, which allows processing of smaller matrices with lower ranks.

We demonstrate the capability of SeNMFk-SPLIT by modeling topics on more than 168,000 abstracts about artificial intelligence (AI) and machine learning (ML) uploaded to arXiv. The distributed library used for SeNMFk-SPLITT is publicly available\footnote{pyDNMFk: \url{https://github.com/lanl/pyDNMFk}} \cite{pyDNMFk}.

\section{Relevant Work}
\label{sec:relevant_work}
Topic modeling provides a convenient way to analyze large amounts of text data. The goal is to segment set of documents on the basis of structural similarities between them. Commonly topic modeling techniques include Latent Semantic Analysis (LSA) \cite{lsa}, Probabilistic Latent Semantic Analysis (PLSA) \cite{plsa}, Latent Dirichlet Allocation (LDA) \cite{lda}, and Non-negative Matrix Factorization (NMF) \cite{xu2003document}.

Bastani et al. used LDA to organize and analyze complaints filed with the Consumer Financial Protection Bureau \cite{BASTANI2019256}. Hazen applied PLSA to a dataset of transcribed telephone calls in an attempt to categorize and summarize conversations \cite{hazen2011latent}. Most similar to this paper's dataset of choice, Rosca et al. used LDA on a collection of documents on AI-assisted legal research \cite{Rosca2020ReturnOT}. Rosca et al. manually determine 35 topics with the help of subject-matter-experts and make a conjecture on the evolution of AI-assisted legal research over the last 50 years. While PLSA and LDA are both probabilistic methods, NMF employs low-rank approximation where the matrix-based representation of the text is factorized to retrieve the manually selected $k$ number of latent topics.

SeNMFk selects the number of latent topics $k$ based on the stability of the extracted topics and includes semantics for the sake of extracting high-quality topics \cite{9521777}. SeNMFk identifies the number of latent topics (automatic model selection) by analyzing sets of NMF minimizations via custom clustering and Silhouette statistics, used to estimate the robustness and accuracy of multiple NMF solutions for different values of the latent variable $k$ \cite{nebgen2021neural}. In addition to the automatic model selection, SeNMFk incorporates semantic information from the text by jointly factorizing the term-document matrix $\mat{X}$ with word-context matrix $\mat{M} \in \rm I\!R_{+}^{m \times m}$. Joint factorization yields more coherent, high quality topics \cite{vangara2020identification}. However, the task of the joint factorization results in an increased memory and computational complexity, especially for a large number of topics and noisy data. In this work we introduce our method SPLIT for SeNMFk solutions. SPLIT enables joint factorization of large corpora by decomposing the word-context ($\mat{M}$) and term-document ($\mat{X})$ matrices separately first, then performing joint factorization of the resulting latent factors to find the common topics, and finally combines the latent information. This allows our approach to work on smaller matrices, breaking down the space complexity to separate operations.

\section{Dataset and Pre-processing}
\label{sec:dataset}

We apply SeNMFk-SPLIT on the abstracts of AI/ML scientific literature uploaded to arXiv\footnote{arXiv Dataset: \url{https://www.kaggle.com/datasets/Cornell-University/arxiv}} \cite{clement2019arxiv}. We focus on papers that are self-reported categories, such as \textit{cs.AI} (Artificial Intelligence), \textit{cs.LG} (Machine Learning), \textit{cs.CL} (Computation and Language), \textit{cs.NE} (Neural and Evolutionary Computing), and \textit{cs.MA} (Multiagent Systems). Our pre-processing procedure includes removal of common stop-words, symbols and next-line characters, non-ASCII characters, e-mail addresses, digits, and tags. We also apply lemmatization using the Python package \textit{NLTK} \cite{bird2009natural}, and exclude non-English abstracts using the Python implementation of \textit{language-detection} software \cite{nakatani2010langdetect}. After removing the abstracts with less than 20 tokens, we obtain 168,177 documents in our corpus. Finally, when selecting the vocabulary we exclude tokens that appeared in less than five documents or in more than 50\% of the corpus, further removing noise. After pre-processing, our final vocabulary includes 25,869 unique terms. We represent our corpus with a TF-IDF matrix $\mat{X} \in \rm I\!R_{+}^{25,869 \times 168,177}$. The semantics of the text are represented with the word-context/co-occurrence matrix $\mat{M} \in \rm I\!R_{+}^{25,869 \times 25,869}$ where the values represent the number of times two words co-occur in a predetermined window length of $w=100$ tokens. We normalize $\mat{M}$ with Shifted Positive Point-wise Mutual Information (SPPMI) \cite{levy2014neural}, with shift $s=4$.

\section{SeNMFk-SPLIT}
\label{sec:methods}

SeNMFk is a NMF method that automatically determines the number of latent topics and extracts coherent topics by exploiting the semantic representation encoded in the word-context matrix adjoined to the TF-IDF. Given the TF-IDF matrix  $\mat{X}$ and normalized word-context SPPMI matrix $\mat{M}$, SeNMFk extracts the topics - the columns of the matrix $\mat{W}$, and the coordinates of the documents - the columns of the matrix $\mat{H}$, as well as a secondary mixing matrix $\mat{G}$.
SeNMFk is performed by solving the joint optimization problem:
\begin{equation}
\underset{\mat{W} \in \mathbb{R}_{+}^{F\times k}, \mat{H} \in \mathbb{R}_{+}^{k \times N}, \mat{G} \in \mathbb{R}_{+}^{k \times F}}{\operatorname{minimize}}
\frac{1}{2} ||\mat{X} - \mat{WH}||_F^2 + \alpha ||\mat{M} - \mat{WG}||_F^2
\end{equation}

where $||.||_F^2$ is the Frobenius norm, and $\alpha$ is a regularization parameter controlling the weight of the semantic SPPMI in the decomposition.
SeNMFk solves the above expression by concatenating the TF-IDF matrix $\mat{X}$ with the SPPMI matrix, $\mat{M}$, and applying pyDNMFk on the concatenation, \cite{stanev2021topic}. 

SeNMFk-SPLIT includes the following optimizations,
\begin{equation}
\label{eqn1}
\mat{W}_1,\mat{H}_1 = \underset{\mat{W}_1 \in \mathbb{R}_{+}^{N\times k_1}, \mat{H}_1 \in \mathbb{R}_{+}^{k_1 \times M}}{\operatorname{minimize}} ||\mat{X} - \mat{W}_1\mat{H}_1||_F^2 
\end{equation}
\begin{equation}
\label{eqn2}
\mat{W}_2,\mat{H}_2 = \underset{\mat{W}_2 \in \mathbb{R}_{+}^{N\times k_2}, \mat{H}_2 \in \mathbb{R}_{+}^{k_2 \times N}}{\operatorname{minimize}} ||\mat{M} - \mat{W}_2\mat{H}_2||_F^2
\end{equation}

Finally,
\begin{equation}
\label{eqn3}
\mat{W},\mat{H}^* = \underset{W \in \mathbb{R}_{+}^{N\times k}, \mat{H}^* \in \mathbb{R}_{+}^{k \times (k_1+k_2)}}{\operatorname{minimize}} ||[\mat{W}_1|\mat{W}_2] - \mat{W}\mat{H}^*||_F^2
\end{equation}
First, SeNMFk-SPLIT estimates the number of latent features $k_1$ and $k_2$ in Equations ~\ref{eqn1}, \ref{eqn2}, and extracts the factor matrices [$\mat{W}_1$ ,$\mat{H}_1$], and [$\mat{W}_2$, $\mat{H}_2$] via pyDNMFk. Next, it concatenates the normalized topic matrices $\mat{W}_1$ and $\mat{W}_2$ to find the common $k$ factors $(k \leq (k_1+k_2))$, and to avoid taking into account factors that are co-linear, or/and linear combinations of other factors. To do that, the concatenated matrix $[\mat{W}_1|\mat{W}_2]$ is factorized by pyDNMFk,  which results in a new topic matrix $\mat{W}$, with common latent topics, and the corresponding mixing matrix $\mat{H}^*$, Equation \ref{eqn3}. The final decomposition of $\mat{X}$ is achieved by learning the coordinates of the documents in $\mat{X}$ in the space of the new common topics through a regression,
\begin{equation}
\label{eqn5}
\mat{H} =  \underset{\mat{H} \in \mathbb{R}_{+}^{k \times M}}{\operatorname{minimize}} ||\mat{X} - \mat{W}\mat{H}||_F^2
\end{equation}

To obtain the final clusters of documents, we assign each document to a given topic via the \emph{argmax} operation along the columns of the matrix $\mat{H}$. \emph{Argmax} is a common operation for probabilistic multi-class classification. Since the columns of $\mat{H}$ are the coordinates of the documents in the space of the latent topics, here this assignment is equivalent to a soft clustering of the documents \cite{vangara2021finding}.

\section{Results}
\label{sec:results}
\begin{figure}[htb]
  \centering
  \includegraphics[width=.95\linewidth]{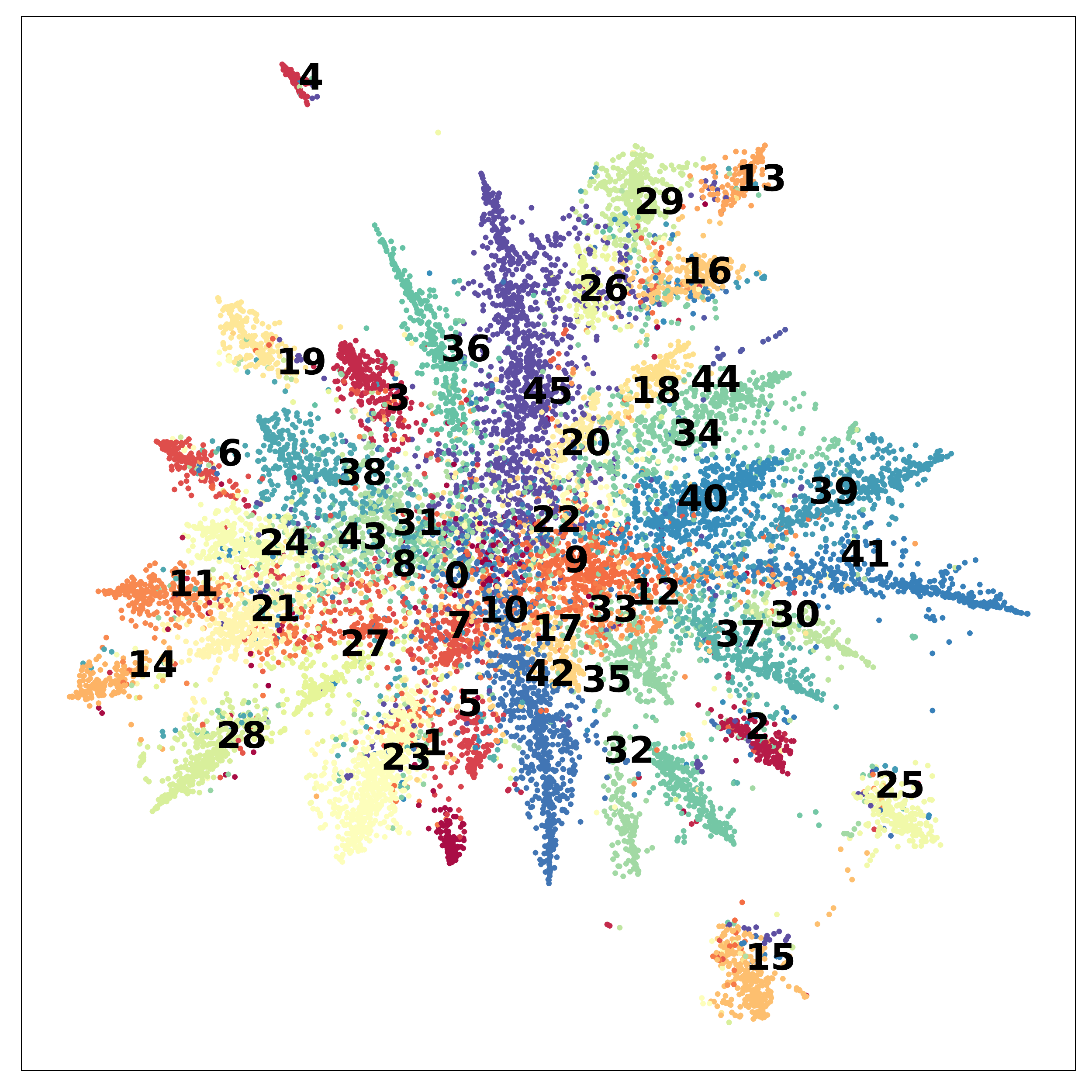}
  \caption{UMAP of $\mat{H}$-clustering, where topic numbers are placed at cluster centroids, and color of each paper represents the topic assignments. \label{fig:umap_plot}}
 % \Description{}
\end{figure}

\begin{figure}[htb]
\centering
\begin{minipage}[b]{.162\textwidth}
\includegraphics[width=\textwidth]{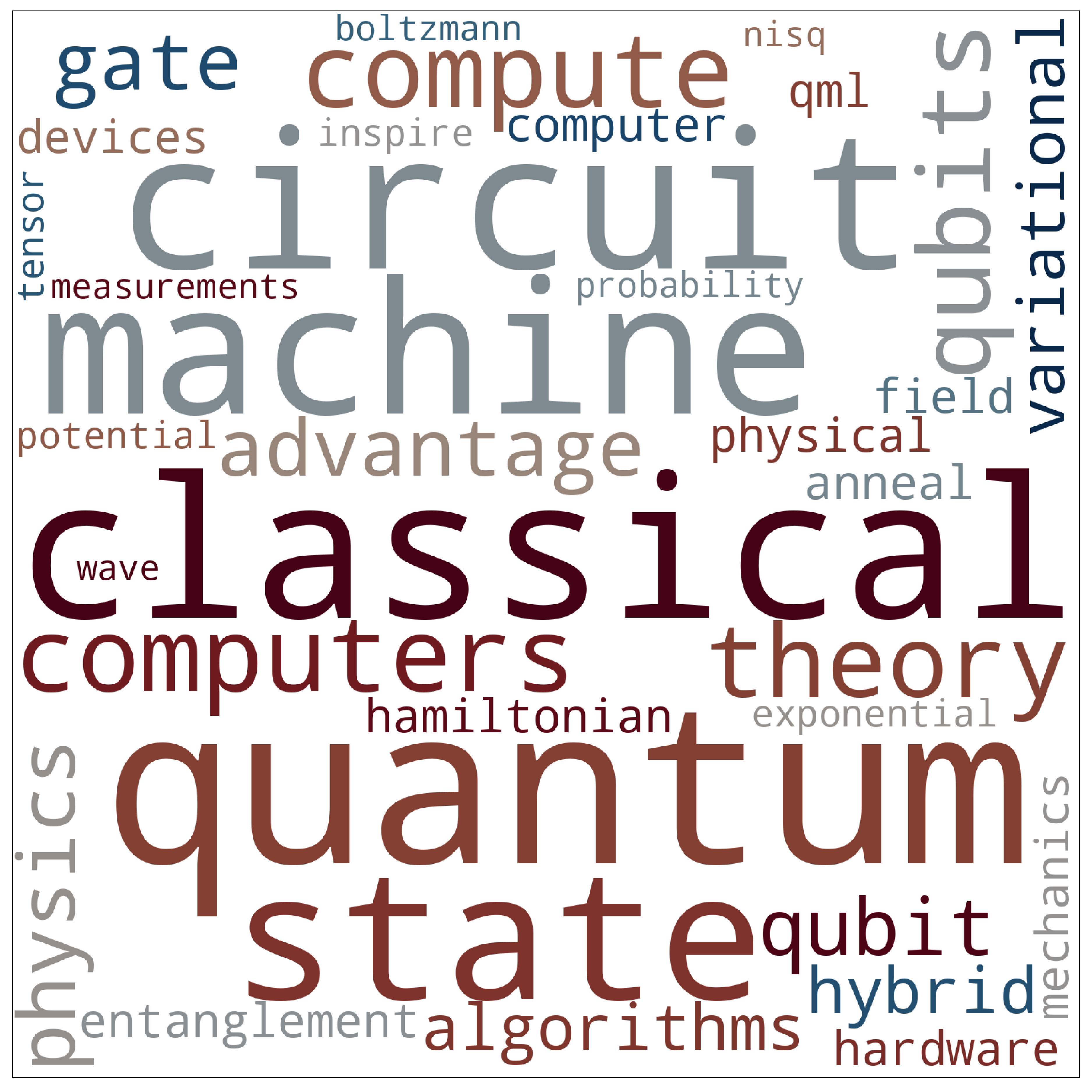}
%\captionsetup{justification=centering}
\caption{Quantum Computing (4)}\label{fig:c4_wc}
\end{minipage}\qquad
\begin{minipage}[b]{.162\textwidth}
\includegraphics[width=\textwidth]{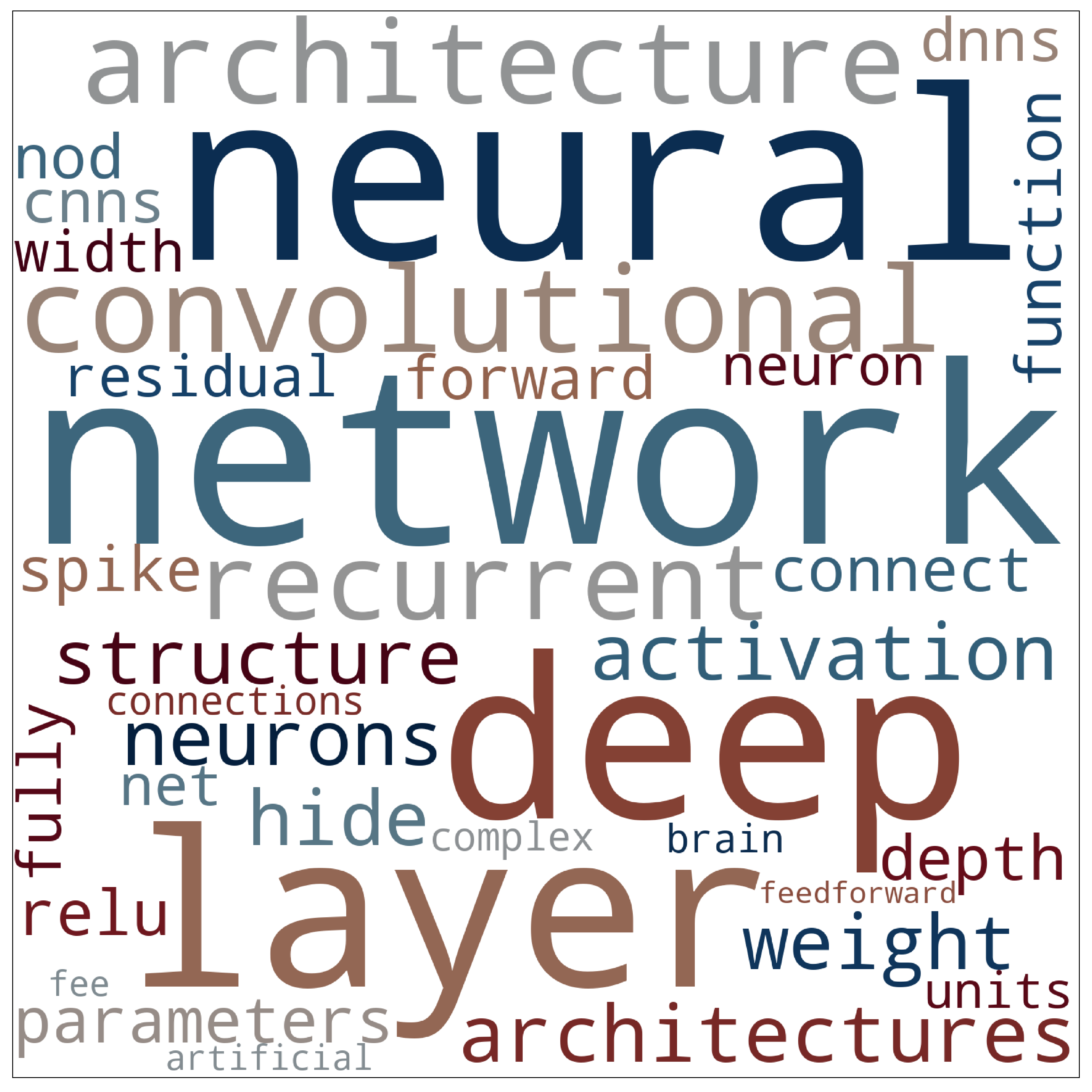}
\caption{Deep Learning (9)}\label{fig:c9_wc}
\end{minipage}
\end{figure}
\begin{figure}[htb]
\centering
\begin{minipage}[b]{.162\textwidth}
\includegraphics[width=\textwidth]{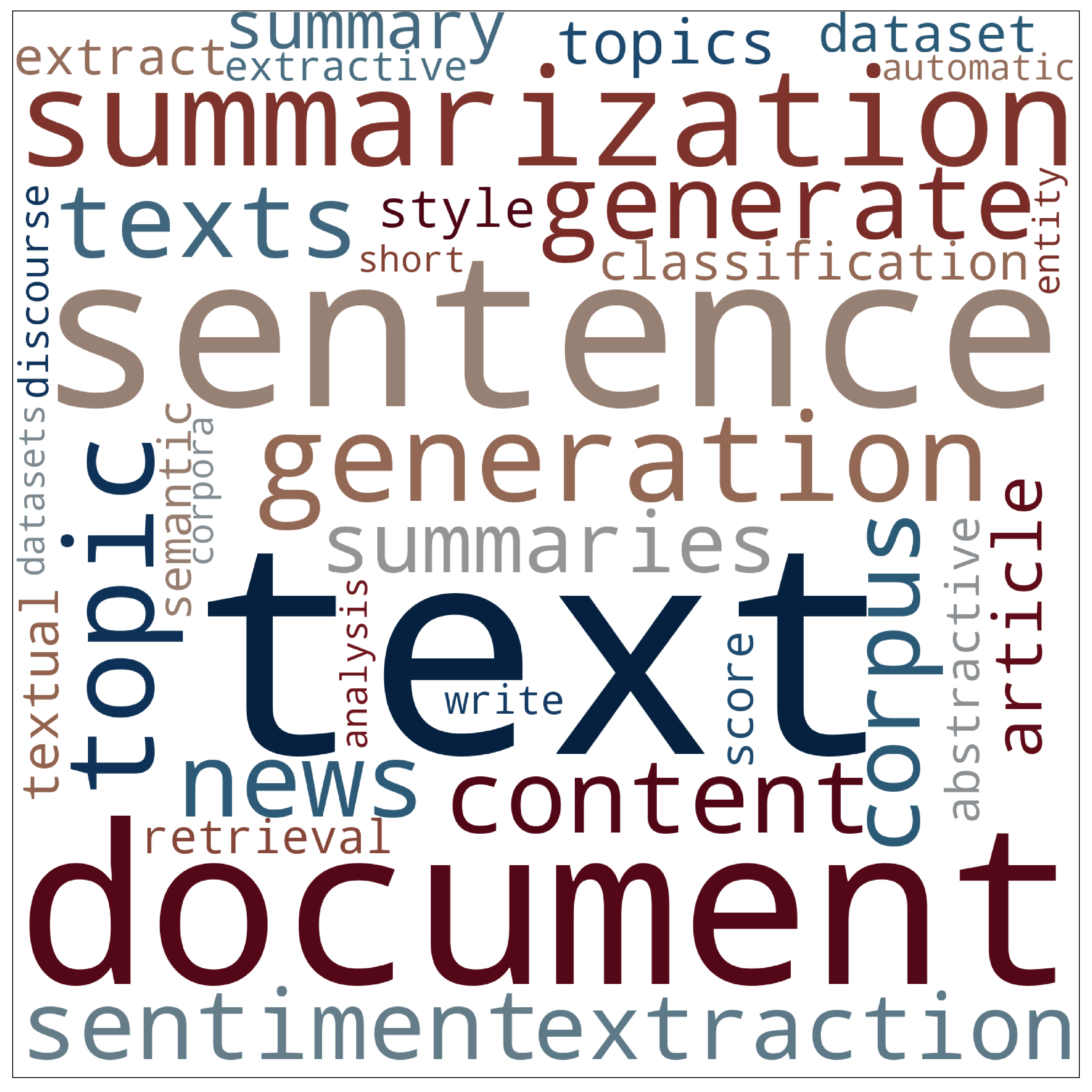}
%\captionsetup{justification=centering}
\caption{Document Engineering (24)}\label{fig:c24_wc}
\end{minipage}\qquad
\begin{minipage}[b]{.162\textwidth}
\includegraphics[width=\textwidth]{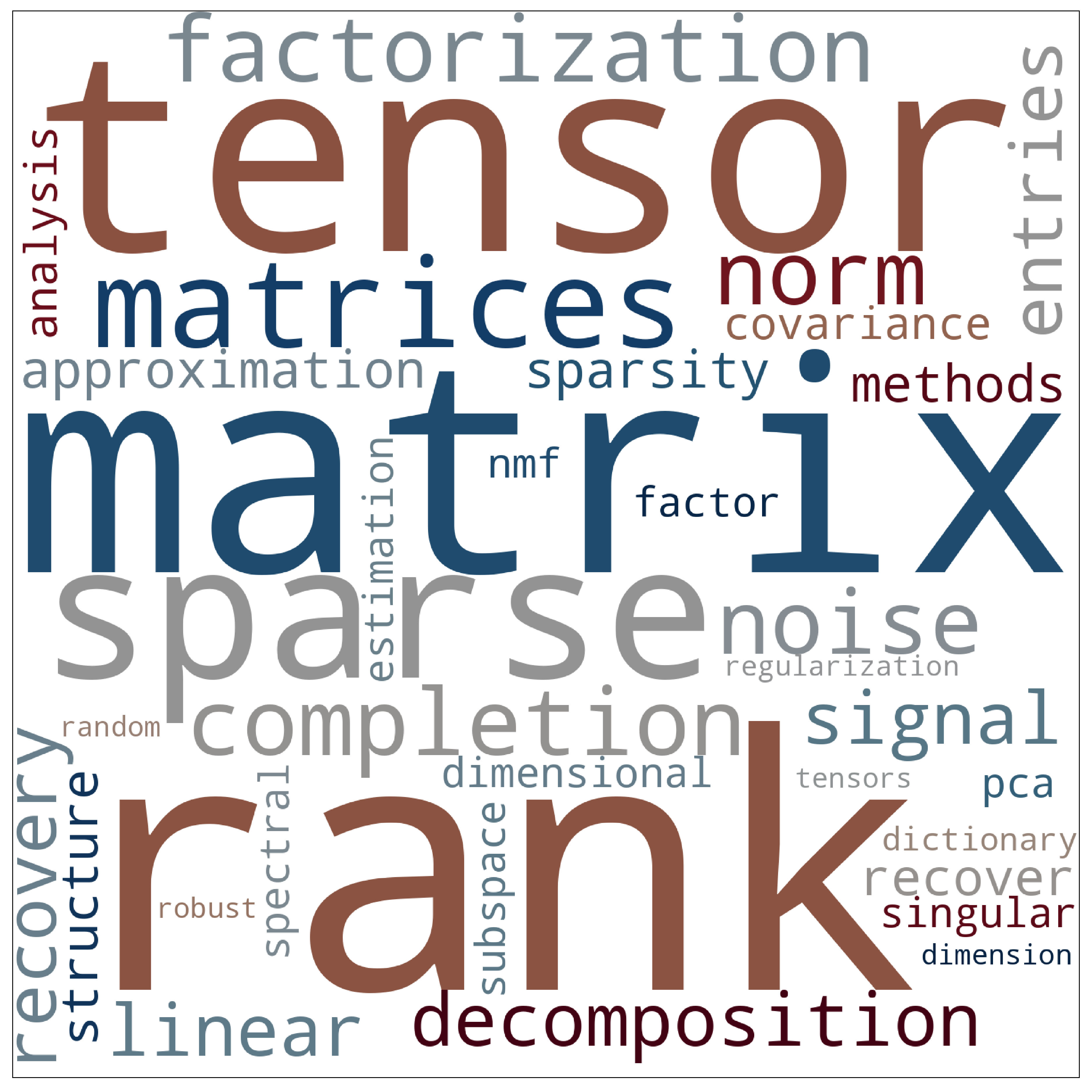}
\caption{Tensor Factorization (37)}\label{fig:c37_wc}
\end{minipage}
\end{figure}

\begin{figure}[htb]
\centering
\includegraphics[width=1\columnwidth]{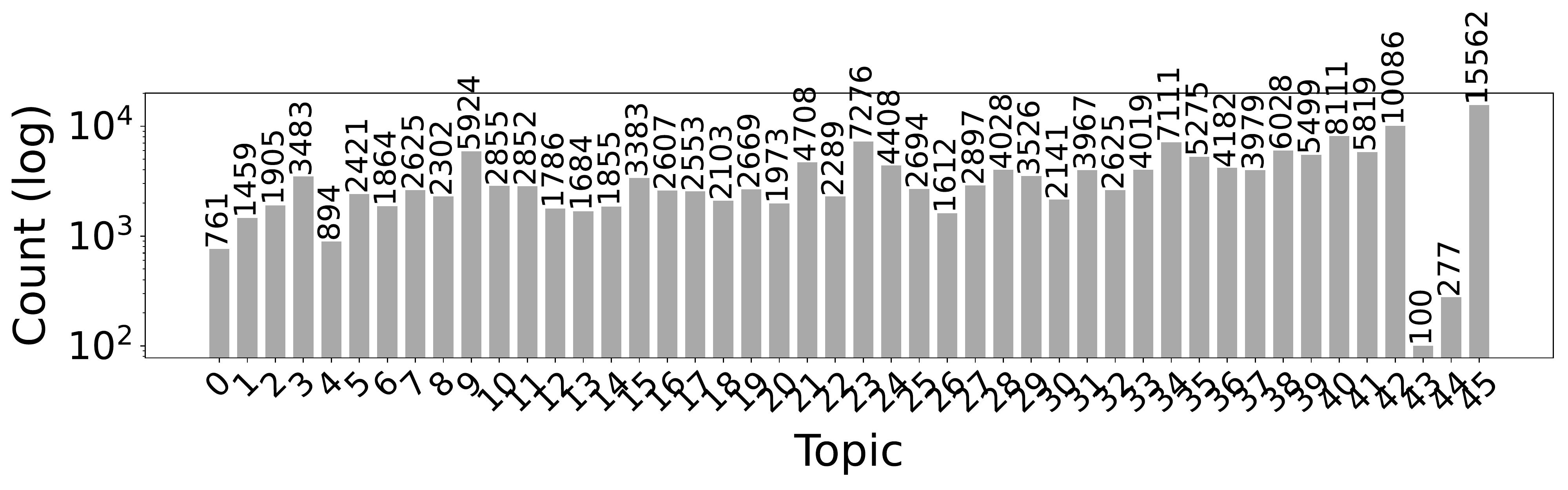}
\vspace*{-4mm}
\caption{Distribution of the abstracts for the detected topics. \label{fig:topic_stats}}
\vspace*{-5mm}
\end{figure}

\begin{table}[htb]
\centering
\caption{Top 5 words in each topic, and the themes we have assigned to these topics based the top 20 words.}
\label{fig:topics_table}
\adjustbox{max width=\columnwidth}{
\begin{tabular}{rll}
%\toprule
 \textbf{Topic} &                                      \textbf{Top 5 Words} &                            \textbf{Theme} \\
%\midrule
0 &   neural, datasets, performance, generative, model &            Neural Network Theory \\
     1 &      transfer, source, domains, domain, adaptation &         Knowledge Generalization \\
     2 &  unsupervised, spectral, algorithms, mean, cluster &            Unsupervised Learning \\
     3 &                  structure, graph, nod, gnns, node &                     Graph Theory \\
     4 &        state, quantum, classical, machine, circuit &                Quantum Computing \\
     5 &           noisy, semi, unlabeled, supervise, label &        Weakly Supervised Systems \\
     6 &               vqa, reason, question, query, answer &           Visual Query Answering \\
     7 & methods, classification, extract, selection, fe... &               Feature Extraction \\
     8 &           performance, multi, transfer, task, meta &                    Meta-learning \\
     9 &        convolutional, layer, network, deep, neural &                    Deep Learning \\
    10 &                fine, tune, performance, pre, train &              Model Tuning \\
    11 &        sentence, semantic, embed, embeddings, word &                   Word Embedding \\
    12 & distributions, distribution, sample, generate, ... &               Generative Systems \\
    13 &           equilibrium, play, players, player, game &                      Game Theory \\
    14 &       sentence, english, translation, nmt, machine &       Neural Machine Translation \\
    15 & perturbations, examples, robustness, attack, ad... &              Adversarial Attacks \\
    16 &    action, policies, reward, policy, reinforcement &           Reinforcement Learning \\
    17 & loss, classifiers, classifier, class, classific... &           Classification Systems \\
    18 &            space, architecture, query, nas, search &       Neural Architecture Search \\
    19 &                track, scene, video, visual, object &                  Computer Vision \\
    20 &              rule, make, structure, decision, tree & Decision Trees and Random Forest \\
    21 &    nlp, multilingual, natural, languages, language &      Natural Language Processing \\
    22 &               instance, consider, rule, point, set &              Rule based Systems \\
    23 &      caption, visual, medical, segmentation, image &                  Computer Vision \\
    24 & generation, sentence, document, summarization, ... &             Document Engineering \\
    25 &  clients, differential, private, federate, privacy &                     Data Privacy \\
    26 &                  path, robot, action, motion, plan &                    Robot Path Planning \\
    27 &  mechanism, self, attention, sequence, transformer &             Attention Mechanisms \\
    28 &             end, asr, speaker, recognition, speech &               Speech Recognition \\
    29 &   communication, agent, environment, agents, multi &              Multi-agent Systems \\
    30 &      gaussian, linear, regression, kernel, kernels &                   Kernel Methods \\
    31 &  relation, entities, semantic, information, entity &               Information Theory \\
    32 &                gender, fair, group, fairness, bias &                   Fairness in AI \\
    33 & generative, representation, representations, la... &          Representation Learning \\
    34 & objective, solve, algorithms, problems, optimize... &                Optimization problems \\
    35 &           items, recommendation, item, users, user &           Recommendation Systems \\
    36 &            semantics, reason, program, logic, rule &                Reasoning Systems \\
    37 &             matrices, rank, matrix, tensor, sparse &            Tensor Factorizations \\
    38 &  evaluation, dialogue, generate, human, generation &                 Dialogue Systems \\
    39 &    sgd, descent, gradient, stochastic, convergence &               Optimizing Systems \\
    40 & distribution, uncertainty, causal, inference, b... &                        Casual AI \\
    41 &                   bandit, bind, arm, bound, regret &              Multi-armed Bandits \\
    42 &    predict, temporal, prediction, series, forecast &             Predictive Analytics \\
    43 & schneider, snacs, supersenses, adpositions, adp... &              Linguistic Analysis \\
    44 &                   gecco, umda, lehre, binval, dang &          Evolutionary Algorithms \\
    45 &          safety, dynamics, drive, systems, control &                  Autonomous Vehicles \\
%\bottomrule
\end{tabular}
}
\vspace*{-5mm}
\end{table}

SeNMFk-SPLIT extracted 46 topics from our AI/ML corpus, where the 20 most prominent words from each topic are examined to determine the themes. The word clouds corresponding to these words are shown for four of the topics in Figures \ref{fig:c4_wc} \ref{fig:c9_wc}, \ref{fig:c24_wc}, and \ref{fig:c37_wc}. We have also listed the top five words and themes of each topic in Table \ref{fig:topics_table}. Word clouds and the top five words demonstrate that each topic features relevant words, highlighting the quality of the topics. In addition, we provide a Uniform Manifold Approximation and Projection (UMAP) \cite{mcinnes2018umap} to visualise the clusters of abstract in Figure \ref{fig:umap_plot}, where the topic number and the color of each document corresponds to different cluster obtained by $\mat{H}$-clustering. 

In Figure \ref{fig:umap_plot}, documents with the same topic are grouped together. Furthermore, the clusters corresponding to related topics seem closer compared to unrelated ones. For example, topics Game Theory \textbf{(13)}, Reinforcement Learning \textbf{(16)}, Multi-agent Systems \textbf{(29)}, Robot Path Planning \textbf{(26)}, and Autonomous Vehicles \textbf{(45)} are related, and the clusters corresponding to these topics are overlapping. Similarly, the clusters corresponding to the topics Neural Machine Translation \textbf{(14)}, Word Embedding \textbf{(11)}, Attention Mechanism \textbf{(27)}, Natural Language Processing \textbf{(21)}, and Document Engineering \textbf{(24)}, are located in close proximity. This relationship can also be explored for the rest of the topics in Figure \ref{fig:umap_plot}. In contrast, the clusters corresponding to the topics Quantum Computing \textbf{(4)}, Data Privacy \textbf{(25)}, and Adversarial Attacks \textbf{(15)} are further away from any other topic clusters as the research corresponding to these topics are emerging fields, and are relatively different directions when compared to the research associated with other topics. Manual investigation of a small set of documents also helps confirm the cogency of the document topic assignment. This demonstrates the ability of the SeNMFk-SPLIT to extract high-quality/interpretable topics. 

Figure \ref{fig:topic_stats} displays the histogram of document topic assignments. Note the small number of documents published in the research topics Linguistic Analysis \textbf{(43)} and Evolutionary Algorithms \textbf{(44)} from Table~\ref{fig:topics_table}, while a significant number of documents belong to the topics Computer Vision \textbf{(23)} and Autonomous Vehicles \textbf{(45)}. These statistics concur with the evolving research in the field of AI.

\section{Conclusion}
\label{sec:conclusion}
In this paper we have introduced SeNMFk-SPLIT, a new topic modeling method that performs automatic estimation of the number of topics and incorporates semantics in modeling. Our method expands upon SeNMFk with a design that can be utilized on large datasets. We demonstrated the capability of SeNMFk-SPLIT on a large corpus containing 168k+ abstracts from AI/ML scientific literature, and showed that our method extracts high-quality topics.

\section{Acknowledgements}
  This manuscript has been approved for unlimited release and has been assigned LA-UR-22-26571. This research was funded by the Los Alamos National Laboratory (LANL) Laboratory Directed Research and Development (LDRD) grant 20190020DR and the LANL Institutional Computing Program, supported by the U.S. Department of Energy National Nuclear Security Administration under Contract No. 89233218CNA000001.

%%
%% The next two lines define the bibliography style to be used, and
%% the bibliography file.
\bibliographystyle{plain}
\bibliography{main}

\end{document}